\documentclass[twocolumn,showpacs,preprintnumbers,amsmath,amssymb]{revtex4-1}
\usepackage{graphicx}
\usepackage{textcomp}

\begin{document}
%\draft
\title{Non-Hermitian tight-binding network engineering}
  \normalsize
\author{Stefano Longhi}
\address{Dipartimento di Fisica, Politecnico di Milano and Istituto di Fotonica e Nanotecnologie del Consiglio Nazionale delle Ricerche, Piazza L. da Vinci
32, I-20133 Milano, Italy}

%\date{.}
%
\bigskip
\begin{abstract}
We suggest a simple method  to engineer a tight-binding quantum network based on proper coupling to an auxiliary non-Hermitian cluster. In particular, it is shown that effective complex non-Hermitian hopping rates can be realized with only complex on-site energies in the network. Three applications of the Hamiltonian engineering  method are presented: the synthesis of a nearly transparent defect in an Hermitian linear lattice; the realization of the Fano-Anderson model with complex coupling; and the synthesis of a $\mathcal{PT}$-symmetric tight-binding lattice with a bound state in the continuum.
\noindent

\end{abstract}


% 03.65.-w Quantum mechanics
% 72.20.Ee    Mobility edges; hopping transport
% 42.82.Et  Waveguides, couplers, and arrays 
% 11.30.Er Charge conjugation, parity, time reversal, and other discrete symmetries
% Supersymmetry,

\maketitle

\section{Introduction}

Hamiltonian engineering is a powerful technique to control classical and quantum phenomena with important applications in many areas of physics such as quantum control \cite{r1,r2,r3,r3bis}, quantum state transfer and quantum information processing \cite{r4,r5,r6,r7,r7b,r8}, quantum simulation \cite{r9,r9b,r9c}, and topological phases of matter \cite{r10,r11,r12,r13,r14,r15}. In quantum systems described by a tight-binding Hamiltonian, quantum engineering is usually aimed at tailoring and controlling hopping rates and site energies, using either static or dynamic methods. For example, special tailoring of the hopping rates in a linear tight-binding chain allows one to realize perfect state transfer between distant sites in the chain \cite{r5,r7b,r16}, whereas external time-dependent perturbations represent a rich and versatile resource to realize synthetic gauge fields, thus achieving topological phases in systems that are topologically trivial in equilibrium \cite{r11,r12,r13,r14,r15}.  The recent growing interest in non-Hermitian quantum and classical systems \cite{r17}, especially in those possessing $\mathcal{PT}$ symmetry \cite{r18}, has motivated the extension of quantum control  methods and Hamiltonian engineering  into the non-Hermitian realm \cite{r18b,r18c,r19,r20,r21}, with ramifications and important applications to  e.g. $\mathcal{PT}$-symmetric integrated photonic devices \cite{r22}.  The ability to tailor complex on-site potentials and non-Hermitian hopping amplitudes is a key task in the engineering of non-Hermitian quantum networks \cite{r18b}. While the engineering of complex on-site potentials is a rather feasible task, 
the realization of complex hopping amplitudes remains a rather challenging issue. For example, in optics non-Hermitian tight-binding networks with complex on-site potentials are readily implemented by evanescent coupling of light modes trapped in optical waveguides or resonators with optical gain and loss in them, while the realization of controllable non-Hermitian coupling constants is a much less trivial task. However, complex hopping amplitudes play an important role for the observation of a wide variety of phenomena that have been disclosed in recent works \cite{r18c,r23,r24,r25,r26,r27,r28}. These include incoherent control of non-Hermitian Bose-Hubbard dimers \cite{r18c}, self-sustained emission in semi-infinite non-Hermitian systems at the exceptional point \cite{r23}, optical simulation of $\mathcal{PT}$-symmetric quantum field theories in the ghost regime \cite{r24,r25}, invisible defects  in tight-binding lattices \cite{r26}, non-Hermitian bound states in the continuum \cite{r27}, and Bloch oscillations with trajectories in complex plane \cite{r28}. Previous proposals to implement complex hopping amplitudes are based on fast temporal modulations of complex on-site energies \cite{r25,r26}, however such methods are rather challenging in practice and, as a matter of fact, to date there is not any experimental demonstration of non-Hermitian complex couplings in tight-binding networks.\\

In this work we suggest a simple method to engineer hopping amplitudes and site energies in a tight-binding network, which simply involves Hermitian couplings and no synthetic gauge fields.  The method is based on proper coupling of the main tight-binding network to an auxiliary non-Hermitian cluster. In particular, it is shown that effective complex (non-Hermitian) hopping rates can be realized with only {\it static} on-site complex potentials in the network, i.e. avoiding fast modulation and thus greatly simplifying its practical implementation. Three applications of the tight-binding network engineering  method are presented: the synthesis of a nearly invisible defect in an Hermitian tight-binding linear lattice; the realization of the Fano-Anderson model with complex coupling; and the synthesis of a $\mathcal{PT}$-symmetric tight-binding lattice with a bound state in the continuum.

\section{Network engineering method}
Let us consider a rather general tight-binding network S, which is constructed topologically by $N$ sites $|n \rangle_S$ and the various connections between them. As a simplified model, it captures the essential features of many discrete classical and quantum systems \cite{r18b}. To engineer the hopping rates and site potentials of the network S, we consider an auxiliary cluster A, with $M$ sites $| \alpha \rangle_A$, which is coupled to the main network S [Fig.1(a)]. As a limiting case, the network S can comprise an infinite number of sites, for example it can describe an infinitely-extended one-dimensional tight-binding lattice, side coupled to the auxiliary cluster A. The tight-binding Hamiltonian of the full system S+A is given by
\begin{equation}
\hat{H}= \hat{H}_S+\hat{H}_A+ \hat{H}_I
\end{equation}
where
\begin{equation}
\hat{H}_S=\sum_{n,m} \mathcal{H}^{(S)}_{n,m} | n \rangle_S \langle m|_S \; ,\;\;  \hat{H}_A=\sum_{\alpha,\beta=1}^M \mathcal{H}^{(A)}_{\alpha,\beta} | \alpha \rangle_A \langle \beta |_A
\end{equation}
are the Hamiltonians of the main (S) and auxiliary (A) networks, respectively, and
\begin{equation}
\hat{H}_I= \sum_{ \alpha=1}^{M} \sum_{n}  \left( \rho_{\alpha,n} |\alpha \rangle_A \langle n |_S +\tilde{\rho}_{n,\alpha} | n \rangle _S \langle \alpha|_A \right)
\end{equation}
describes their interaction. In the above equations, the roman and greek indices run over the sites of main network S and auxiliary cluster A, respectively,
the matrix $\mathcal{H}^{(S)}$ describes on-site potentials (diagonal elements $\mathcal{H}^{(S)}_{n,n}$) and hopping amplitudes (off-diagonal elements $\mathcal{H}^{(S)}_{n,m}$, $ n \neq m$) among the various sites of the main system S, and the $M \times M$ matrix $\mathcal{H}^{(A)}$ is the analogous matrix for the auxiliary system. The two
matrices $\rho$ and $\tilde {\rho}$, entering in Eq.(3),
describe the interaction between the sites of S and A.  We assume that the hopping rates among the different sites in both main (S) and auxiliary (A) networks are Hermitian and that there are not gauge fields that introduce Peierls' phases in the hopping amplitudes. Such an assumption implies that the non-diagonal elements of the matrices $\mathcal{H}^{(S)}$ and $\mathcal{H}^{(A)}$, and all the elements of the matrices $\rho$ and $\tilde{\rho}$, are real, with $\rho_{\alpha,n}= \tilde{\rho}_{n, \alpha}$, $\mathcal{H}^{(S)}_{n,m}=\mathcal{H}^{(S)}_{m,n}$ and $\mathcal{H}^{(A)}_{\alpha, \beta}=\mathcal{H}^{(A)}_{\beta,\alpha}$
  i.e.
\begin{equation}
\tilde{\rho}=\rho^T \; , \; \; \mathcal{H}^{(S)}=\mathcal{H}^{(S) \: T} \; , \; \;  \mathcal{H}^{(A)}=\mathcal{H}^{(A) \: T}.
\end{equation}
However, site potentials $\mathcal{H}^{(S)}_{n,n}$ and $\mathcal{H}^{(A)}_{\alpha,\alpha}$ , in either or both the main and auxiliary networks, are allowed to be complex. Note that, if the auxiliary cluster is made of purely dissipative sites, i.e. the imaginary parts of $\mathcal{H}^{(A)}_{\alpha, \alpha}$ are either zero or negative, the eigenvalues of $\mathcal{H}^{(A)}$ have negative (or vanishing) imaginary parts, and secularly growing terms of the auxiliary site amplitudes are avoided in the weak coupling regime $\rho \rightarrow 0$.
After expanding the state vector of the full system as 
\begin{equation}
|\psi (t) \rangle=\sum_{n} c_n(t)  |n \rangle_S + \sum_{\alpha=1}^{M} a_\alpha (t) |\alpha \rangle_A, 
\end{equation}
from the Schr\"{o}dinger equation $i \partial_t | \psi(t) \rangle = \hat{H} | \psi \rangle$  one obtains
\begin{eqnarray}
i \frac{d c_n}{dt} & = & \sum_{m} \mathcal{H}^{(S)}_{n,m} c_m+\sum_{\alpha=1}^{M} \tilde{\rho}_{n,\alpha} a_{\alpha} \\
i \frac{d a_{\alpha}}{dt} & = & \sum_{\beta=1}^M \mathcal{H}^{(A)}_{\alpha,\beta} a_{\beta}+\sum_{n} {\rho}_{\alpha,n} c_{n}. 
\end{eqnarray}
To obtain the dynamical behavior of the system solely, one might try to proceed by elimination of the auxiliary amplitudes $a_{\alpha}(t)$ from Eqs.(6) and (7), thus obtaining coupled integro-differential equations for the system amplitudes $c_n(t)$ (see Appendix A).  However, such a procedure turns out to be useful in defining an effective energy-independent Hamiltonian $\hat{H}_{eff}$ for the system S whenever the auxiliary system A is an almost continuum of states (i.e. $M \rightarrow \infty$) and the S-A coupling is weak. Indeed, this is the usual way to describe metastability of Markovian open quantum systems (see, for instance, \cite{r29,r30,r31}). Here, however, we typically consider a finite (and possibly small) number of auxiliary sites $M$, a typical infinitely-extended system S ($N \rightarrow \infty$), and do not necessarily require the S-A coupling to be weak. In such a case, the reduction procedure of open quantum systems and derivation of an effective Hamiltonian can be applied under certain conditions solely, which are discussed in the Appendix A. For our purposes, we follow here a different strategy. Let us look for an eigenstate of $\hat{H}$ with energy $E$, which can be either a bound state or a scattered state when $N= \infty$. Assuming the dependence $\sim \exp(-iEt)$ for the amplitudes in Eqs.(6) and (7), one obtains 
\begin{eqnarray}
E \mathbf{c} & = & \mathcal{H}^{(S)} \mathbf{c}+ \tilde{\rho} \mathbf{a} \\
E \mathbf{a} & = & \mathcal{H}^{(A)} \mathbf{a}+\rho \mathbf{c},
\end{eqnarray}
 where $\mathbf{c}=(... , c_{-1},c_0,c_1,c_2, ...)^T$ and $\mathbf{a}=(a_1,a_2,a_3,..., a_M)^T$ are the vectors of S and A site amplitudes. After elimination of the amplitudes $\mathbf{a}$, one obtains
\begin{equation}
E \mathbf{c}  =  \mathcal{H}_{eff}(E)  \mathbf{c},
\end{equation}
 where we have set
 \begin{equation}
 \mathcal{H}_{eff}(E) \equiv \mathcal{H}^{(S)}+\tilde{\rho}(E-\mathcal{H}^{(A)})^{-1} \rho.
 \end{equation}
 Equation (11) shows that the effect of the auxiliary cluster A is to renormalize the hopping amplitudes and site potentials of the network S by adding, to the Hamiltonian $\mathcal{H}^{(S)}$, a generally {\it energy-dependent} term [the second term on the right hand side of Eq.(11)]. Such an additional term is analogous to the so-called 'optical potential' found in the effective Hamiltonian description of decaying open quantum systems using the Feshbach's projection operator method \cite{r30}.
 \begin{figure}
\includegraphics[scale=0.24]{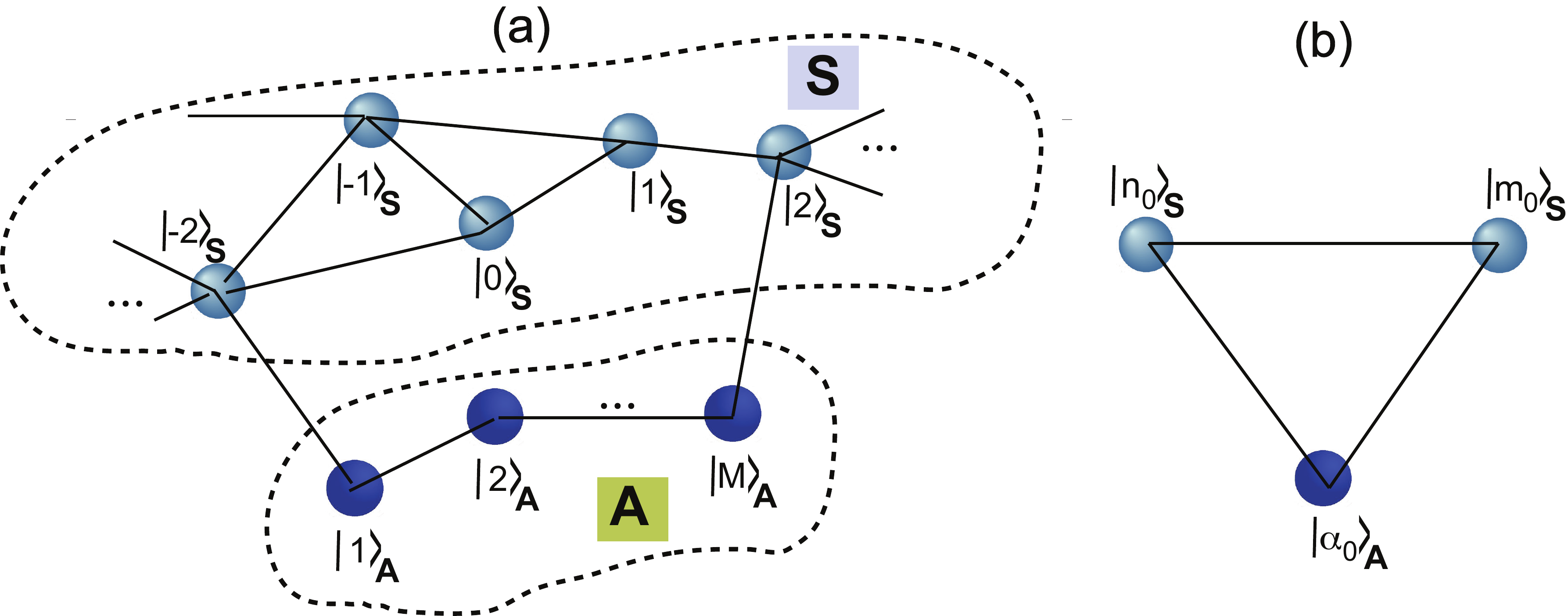}
\caption{(Color online) (a) Schematic of a tight-binding network S coupled to an auxiliary cluster A made of $M$ sites. (b) Control of hopping amplitude between sites $|n_0 \rangle_S$ and $|m_0 \rangle_S$ via an auxiliary site $| \alpha_0 \rangle_A$.}
\end{figure}

 Interestingly, the optical potential term generally makes the off-diagonal elements of $\mathcal{H}_{eff}$ complex, i.e. the effective hopping amplitudes are non-Hermitian in spite all hopping amplitudes in the network and auxiliary cluster are Hermitian. In particular, one can readily shown that:\\
 (i) If  the cluster A is Hermitian,  i.e. on-site potentials $\mathcal{H}^{(A)}_{\alpha,\alpha}$ are real, then $\mathcal{H}_{eff}$ is real and symmetric. In this case the effect of A is to renormalize the hopping rates of sites in S because of additional tunneling paths introduced by the the auxiliary sites, however they remain Hermitian. A typical example is provided by indirect (second-order) tunneling, which is described in Sec.III.A.\\
 (ii) If the on-site potentials $\mathcal{H}^{(A)}_{\alpha,\alpha}$ in the auxiliary sites are complex, for any real energy $E$  the matrix $\mathcal{H}_{eff}(E)$ is symmetric but {\it not} Hermitian, i.e. one has
 \begin{equation}
 (\mathcal{H}_{eff } )_{n,m}= (\mathcal{H}_{eff } )_{m,n}
  \end{equation}
 but $ (\mathcal{H}_{eff } )_{n,m}$ is generally complex. This is one of the most important result of the analysis and shows that complex on-site potentials in the auxiliary sites result in an effective {\it non-Hermitian} hopping amplitudes among sites in the network S.\par
It should be noted that the above equivalence holds for a prescribed energy $E$, and that Eq.(10) is actually an implicit eigenvalue equation since the effective Hamiltonian $\mathcal{H}_{eff}(E)$ depends on energy $E$ via the 'optical potential' term [the second term on the right hand side of Eq.(11)]. For weak S-A coupling, i.e. for $\rho \rightarrow 0$, an iterative procedure can be  used to solve Eq.(10), while in certain special cases some bound states can be determined in a closed form without resorting to any approximation (see, for example, the model discussed in Sec.III.C). However, there are at least two important cases where the problem is amenable of analytical results, without requiring small interaction limit.\\
(i) {\it  Linear tight-binding homogeneous lattices}. Let us suppose that the network S is an infinitely-extended one-dimensional tight-binding lattice with (asymptotically) homogeneous nearest-neighbor hopping rate $ \kappa$ and uniform site potentials, i.e. $\mathcal{H}^{(S)}_{n,m} \rightarrow \kappa ( \delta_{n,m+1}+\delta_{n,m-1})$ as $n,m \rightarrow \pm \infty$. Since  the auxiliary cluster A couples only with a few sites in S with finite index $n$, the scattering states of S are asymptotically plane waves, $c_n \sim \exp(\pm i qn)$  as $ n \rightarrow \pm \infty$, and their energy is known and given by $E=2 \kappa \cos(q)$, where $-\pi \leq q < \pi$ is the Bloch wave number. Hence, for a fixed value of the wave number $q$, the effective Hamiltonian $\mathcal{H}_{eff}$ is known and scattering states , including reflection/transmission coefficients, can be readily determined by standard methods. This approach can be applied  to the determination of bound states as well, looking at the poles of the spectral transmission. An example is discussed in Sec.III.A.\\
(ii) {\it  Large on-site potentials of auxiliary cluster.}  If the on-site potentials $\mathcal{H}^{(A)}_{\alpha,\alpha}$ of the auxiliary sites are (in modulus) much larger than the energy $|E|$, the inverse matrix $(E-\mathcal{H}^{(A)})^{-1}$ entering in the 'optical potential' is weakly dependent on the energy $E$, and thus one can approximately set
\begin{equation} 
 \mathcal{H}_{eff}  \simeq \mathcal{H}^{(S)}- \tilde{\rho} (\mathcal{H}^{(A)})^{-1} \rho.
\end{equation}
In this way the dependence of the effective Hamiltonian $\mathcal{H}_{eff}$ on energy is removed. In particular, by further assuming $|\mathcal{H}^{(A)}_{\alpha,\alpha}| \gg |\mathcal{H}^{(A)}_{\beta,\gamma}|$ ($\beta \neq \gamma$), $(\mathcal{H}^{(A)})^{-1}$ is diagonal with elements $1/ \mathcal{H}^{(A)}_{\alpha,\alpha}$, so that taking into account that $\tilde{\rho}= \rho^T$ one has
\begin{equation}
(\mathcal{H}_{eff})_{n,m} \simeq  \mathcal{H}^{(S)}_{n,m}-\sum_ {\alpha=1}^{M} \frac{\rho_{\alpha,n} \rho_{\alpha,m}}{\mathcal{H}^{(A)}_{\alpha,\alpha}}.
 \end{equation}
Equation (14) enables, in principle, to engineer the effective matrix elements $(\mathcal{H}_{eff})_{n,m}$ in a rather flexible and independent way. For example, to engineer the hopping amplitude between two prescribed sites $n=n_0$ and $m=m_0$ of the network S, we can consider an auxiliary site, say $| \alpha_0 \rangle_A$, which is the only site of A coupled to $|n_0  \rangle_S$ and  $|m_0  \rangle_S$ [Fig.1(b)]. From Eq.(14) one then obtains
\begin{equation}
(\mathcal{H}_{eff})_{n_0,m_0} \simeq \mathcal{H}^{(S)}_{n_0,m_0}- \frac{\rho_{\alpha_0,n_0} \rho_{\alpha_0,m_0}}{\mathcal{H}^{(A)}_{\alpha_0,\alpha_0}}.
 \end{equation}
Note that, while the hopping amplitudes  $(\mathcal{H}^{(S)})_{n_0,m_0}$, $\rho_{\alpha_0,n_0}$ and $\rho_{\alpha_0,m_0}$ are real, the on-site potential $\mathcal{H}^{(A)}_{\alpha_0,\alpha_0}$ is complex, so that by a judicious choice of  $(\mathcal{H}^{(A)})_{\alpha_0,\alpha_0}$ and  $\rho_{\alpha_0,n_0} \rho_{\alpha_0,m_0}$ a desired non-Hermitian complex hopping amplitude $(\mathcal{H}_{eff})_{n_0,m_0} $ can be realized. Note that, as opposed to the weak-coupling limit described in the Appendix A, in such a procedure there is no restriction on the magnitude of the S-A coupling $\rho$, so that the correction to the hopping rate provided by the second term on the right hand side of Eq.(15) is not necessarily small.

\section{Applications}
The rather general procedure of network engineering presented in Sec.II is exemplified by considering three applications to some important physical problems, namely the synthesis of a nearly invisible defect in an Hermitian homogeneous lattice, the realization of the Fano-Anderson model with non-Hermitian coupling, and the synthesis of a $\mathcal{PT}$-symmetric tight-binding lattice with a bound state in the continuum.

\begin{figure}
\includegraphics[scale=0.3]{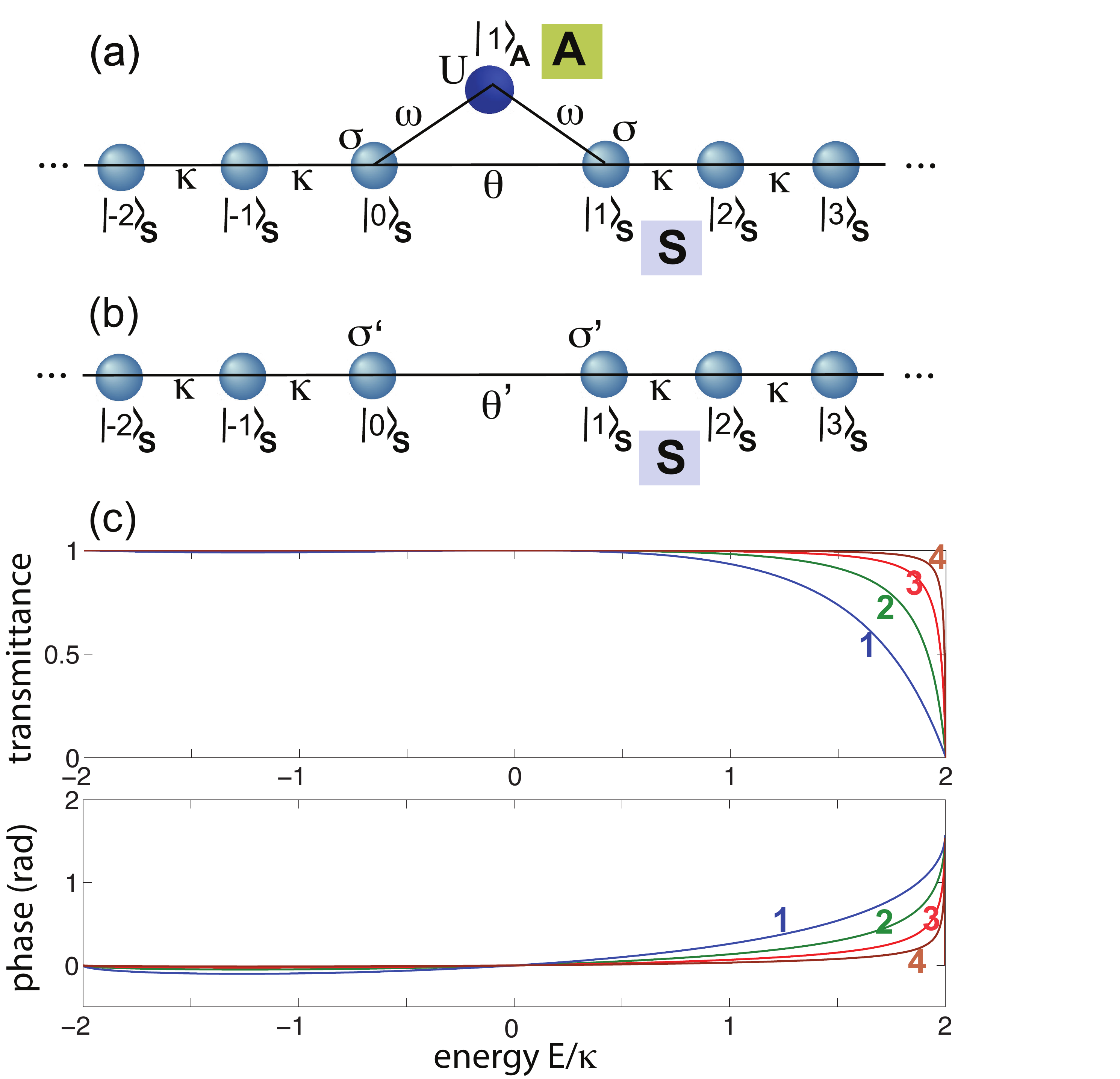}
\caption{(Color online) (a) Schematic of a linear tight-binding lattice S side-coupled to an auxiliary site A. (b) Equivalent lattice after elimination of the auxiliary site, with energy-dependent renormalized hopping rate $\theta'$ and site potential $\sigma'$, defined by Eqs.(20) and (21) given in the text. (c) Behavior of the spectral transmittance $|t(E)|^2$ and phase of $t(E)$ versus energy $E=2 \kappa \cos (q)$ for parameter values $ \theta / \kappa=0.2$, $\sigma/ \kappa=-0.8$ and for: $U/  \kappa=-5$,  $\omega/ \kappa=2$ (curve 1); $U/  \kappa=-10$, $ \omega/ \kappa= 2 \sqrt{2}$ (curve 2);  $U/  \kappa=-20$,  $\omega/ \kappa=4$ (curve 3); $U/  \kappa=-40$,  $\omega/ \kappa=4 \sqrt{2}$ (curve 4).}
\end{figure}

\subsection{Nearly-invisible defect in an Hermitian tight-binding lattice}
The possibility of synthesizing transparent defects in tight-binding lattices has received an increasing interest in the past recent years \cite{r26,r32,r33,r34}, with the experimental demonstration of reflectionless potentials in arrays of evanescently-coupled optical waveguides with tailored coupling constants \cite{r33}.  In optics, reflectionless defects sustaining propagative bound states  offer the possibility to realize transparent optical intersections in photonic circuits \cite{r34}. Transparent defect modes are generally synthesized by inverse scattering or supersymmetric methods \cite{r26,r32}, which require a careful control of hopping amplitudes over several lattice sites. Here it is shown that a nearly invisible defect mode can be simply realized in an otherwise homogeneous tight-binding linear lattice exploiting the hopping rate engineering method discussed in the previous section. Let us consider the linear tight-binding lattice S shown in Fig.2(a), which is side coupled to one auxiliary site A ($M=1$). For such a system we have
\begin{eqnarray}
(\mathcal{H}^{(S)})_{n,m} & =  & \kappa (\delta_{n,m+1}+\delta_{n,m-1})+ \sigma (\delta_{n,0}\delta_{m,0}+\delta_{n,1}\delta_{m,1}) \nonumber \\
& + & (\theta-\kappa)(\delta_{n,0}\delta_{m,1}+\delta_{n,1} \delta_{m,0}) \\
\mathcal{H}^{(A)} & = & U \\
\rho_{1,n} & = & \omega (\delta_{n,0}+\delta_{n,1}).
\end{eqnarray}
 In the above equations, $\kappa$ is the hopping rate between adjacent sites in the main lattice S, with a defective hopping rate $\theta$ between sites $|0 \rangle_S$ and $|1 \rangle_S$; $\sigma$ is the potential at sites $|0 \rangle_S$ and $|1 \rangle_S$; $\omega$ is the hopping rate between the auxiliary site $|1 \rangle_A$ and the two sites $|0 \rangle_S$ and $|1 \rangle_S$; and $U$ is the potential of site $|1 \rangle_A$ [see Fig.2(a)]. According to Eq.(11), elimination of the auxiliary site yields the following effective Hamiltonian for the lattice S
 \begin{equation}
 (\mathcal{H}_{eff})_{n,m}= \mathcal{H}^{(S)}_{n,m}+\frac{\omega^2}{E-U} \sum_{k,l=0}^{1} \delta_{n,l}\delta_{m,k}
 \end{equation}
 which basically describes the modified linear lattice depicted in Fig.2(b). As it can be seen, the role of the side-coupled auxiliary site $|0 \rangle_A$ is to modify the potentials and hopping rate between sites $ | 0 \rangle_{S}$ and  $ | 1 \rangle_{S}$ to the effective values 
 \begin{eqnarray}
 \sigma'=\sigma+\omega^2/(E-U) \\
 \theta'=\theta+\omega^2/(E-U).
 \end{eqnarray}
 In particular, note that the effective hopping rate $\theta'$ is given by the interference of two terms: direct tunneling between sites $|0 \rangle_S$ and $|1 \rangle_S$ with hopping amplitude $\theta$, and indirect (second-order) tunneling via the auxiliary site with energy-dependent hopping amplitude $\omega^2/(E-U)$.
 
  The spectral transmission and reflection of the resulting lattice of Fig.1(b), as well as bound states, can be calculated by standard methods, with the results of the analysis given below. However, the conditions for a nearly-invisible defect can be readily established from an inspection of Eqs.(20) and (21) without any detailed calculation. In fact, as discussed for the general case in Sec.III, in the large $|U|$ limit the effective Hamiltonian turns out to be independent of the energy $E$, the latter being bounded in the interval $(-2 \kappa, 2 \kappa)$ for scattering states. Hence for large $|U|$ one can assume $\sigma' \simeq \sigma-\omega^2/U$ and $\theta' \simeq \theta-\omega^2/U$. Interestingly, with the choice
 \begin{equation}
 \sigma= \theta - \kappa \; , \;\;  \omega^2= U(\theta-\kappa)
 \end{equation} 
one has $\sigma' \simeq 0$ and $\omega' \simeq \kappa$, i.e. the effective lattice in Fig.1(b) is homogeneous and thus invisible. Note that, for $\theta < \kappa$, invisibility is obtained for $U<0$ and $\sigma<0$.
The onset of invisibility can be checked by exact calculation of the spectral transmission and reflection coefficient for the lattice of Fig.1(b) following a standard procedure. Let us look for a scattered state solution to the eigenvalue equation (10) of the form
\begin{equation}
c_n= \left\{
\begin{array}{cc}
\exp(-iqn)+r(q) \exp(iqn) & n \leq 0 \\
t(q) \exp(-iqn) & n \geq 1
\end{array}
\right.
\end{equation} 
where $q$ is the Bloch wave number, $t(q)$ and $r(q)$ are the spectral transmission and reflection coefficients, respectively, and $E=2 \kappa \cos (q)$ is the energy.  The expressions of $t$ and $r$ can be determined by writing coupled equations for amplitudes at sites $|0 \rangle_S$ and $|1 \rangle_S$, i.e.
\begin{eqnarray}
E c_0 & = & \kappa c_{-1}+ \theta' c_1+ \sigma ' c_0 \\
Ec_1 & = & \kappa c_2+ \theta' c_0+ \sigma' c_1.
\end{eqnarray}
After substitution of the Ansatz (23) into Eqs.(24) and (25),  coupled equations for $r$ and $t$ are obtained, which can be solved for $t$ yielding
\begin{equation}
t(q)= \frac{2 i \kappa \theta' \sin (q) \exp(iq)}{[\kappa \exp(iq)-\sigma'+\theta'][\kappa \exp(iq)-\sigma'-\theta']}.
\end{equation}
The dependence of the corrected hopping rate $\theta'$ and site potential $\sigma'$ on energy $E$ is determined by Eqs.(20) and (21). Substitution of Eqs.(20) and (21) into Eq.(26) finally yields
\begin{equation}
t(q)= \frac{2 i\kappa [\theta(E-U)+ \omega^2] \sin(q) \exp(iq)}{[\kappa \exp(iq)-\sigma+\theta]\{ (E-U)[\kappa \exp(iq)-\sigma-\theta]-2 \omega^2\} }.
\end{equation}
A typical behavior of $t(E)$ (modulus and phase) for increasing values of $|U| / \kappa$ and for $\theta= \kappa /5$ are shown in Fig.2(c). The site potential $\sigma$ and hopping rate $\omega$ are chosen to satisfy the invisibility condition Eq.(22). Note that, according to the theoretical prediction, a near invisible defect over the entire tight-binding energy band is realized at increasing values of $|U|/ \kappa$.\\
 Bound states sustained by the lattice at the defective region can be determined by looking at the poles of $t(q)$ in the complex $q$ plane, with $\rm{Im}(q)<0$. Assuming that the conditions (22) are satisfied, after setting $y=\exp(iq)$ the condition $t(q)= \infty$  leads to the following algebraic (cubic) equation for $y$
\begin{equation}
y^3+\left( 1-\frac{U}{\kappa}- \frac{2 \theta}{\kappa}\right) y^2+ \left( 1+\frac{U}{\kappa} \right)y+1-\frac{2 \theta}{\kappa}=0.
\end{equation}
with the constraint $|y|>1$. Assuming $U<0$ and $0 \leq \theta< \kappa$, such an equation admits of one acceptable solution, corresponding to the existence of one bound state, for $U<-2 \theta$. Hence, in the near transparency regime (i.e. for $|U|/ \kappa$ large), there is always one bound state, corresponding in the physical lattice of FIg.1(a) to high localization in the auxiliary site $|1 \rangle_A$.

\begin{figure}
\includegraphics[scale=0.3]{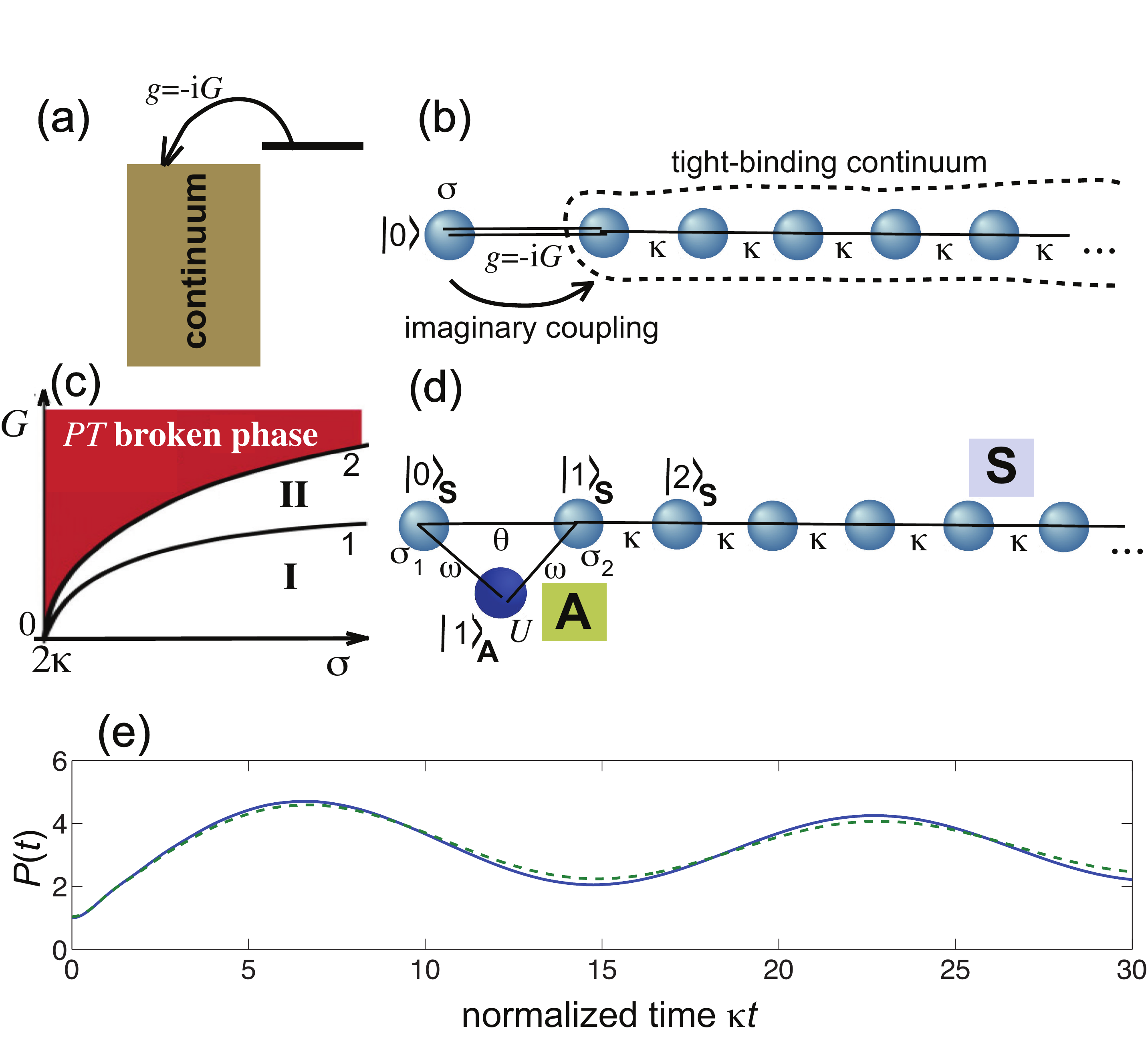}
\caption{(Color online) (a) Schematic of a bound state coupled coupled to a continuum of states. When the energy of the bund state is not embedded in the continuous spectrum, for small coupling $g$ there is not a decay while the energy of the bound state is renormalized. The ghost regime of the Lee model is realized when the coupling $g$ to the continuum is imaginary ($g=-iG$). (b) Tight-binding lattice realization of the Lee model in the ghost regime. The continuum of states is given by the Bloch modes of a semi-infinite homogeneous lattice. The end of the semi-lattice is attached by complex coupling $g$ to the localized state $|0 \rangle$. (c) Phase diagram of the Lee model. Below the curve 2 the system is in the unbroken $\mathcal{PT}$ phase, with one bound state in region I (the "physical" particle state) and two bound states in region II (the "physical" particle state plus the "ghost"). Above curve 2 the $\mathcal{PT}$ symmetry is broken. Analytic equations of curves 1 and 2 are $G=\kappa \sqrt{-2+(\sigma/ \kappa)}$ and $G=\kappa \sqrt{-1+( \sigma / 2 \kappa)^2}$, respectively. (d) Implementation of the non-Hermitian (imaginary) coupling by an auxiliary site $|1 \rangle_A$. (d) Numerically-computed evolution of the occupation probability $P(t)$ as obtained from the exact Lee Hamiltonian (solid curve) and from the synthesized lattice (dashed curve). Parameter values are given in the text.}
\end{figure}

\subsection{Fano-Anderson model with complex coupling}
The Fano-Anderson model \cite{r35,r36}, also referred to as
the Friedrichs-Lee model in quantum field theory \cite{r37}, 
is ubiquitous in different areas of physics and describes quite generally the coupling of a bund
state to a continuum [Fig.3(a)]. A paradigmatic example of the Fano-Anderson model, which is ofter encountered in the theory of coherent transport in mesoscopic condensed-matter systems and in integrated photonic systems,  is provided by the coupling of a localized state to an infinite or semi-infinite tight-binding lattice, i.e. to a continuous band of Bloch modes \cite{r38,r39}. Hermitian coupling (hopping amplitude $g$ real) is the gold standard in such models, however complexification of the coupling constant ($g$ imaginary) is of some interest in certain quantum models, such as the Lee model in the so-called ghost regime \cite{r24} or in the theory of the inverted quantum oscillators and quantum amplifiers \cite{r39b,r39c}. Let us discuss here the Lee model with complex coupling, in which the Hamiltonian is
not Hermitian but $\mathcal{PT}$ symmetric \cite{r24}. A simple tight-binding lattice realization of the Lee model in the ghost regime is shown in Fg.3(b) \cite{r25}. It consists of a semi-infinite homogeneous tight-binding chain with (Hermitian) hopping amplitude $\kappa$ between adjacent sites and connected to the end to a node with on-site potential $\sigma > 2 \kappa$ and with complex coupling $g=-iG$ . The Hamiltonian of the tight-binding Lee model of Fig.3(b) reads
\begin{eqnarray}
\hat{H} & = & \sum_{n=1}^{\infty} \kappa \left ( |n \rangle \langle n+1|+ | n+1 \rangle \langle n | \right)+ \nonumber \\
& + & \sigma |0 \rangle \langle 0| +g(|0 \rangle \langle 1|+|1 \rangle \langle 0|)
\end{eqnarray}
where $g=-iG$ is the imaging coupling of the localized state at site $|0 \rangle$ with the semi-infinite tight-binding lattice.
The phase space diagram of the Lee Hamiltonian (29) is depicted in Fig.3(c) \cite{r25}. In the unbroken $\mathcal{PT}$ phase, the lattice can sustain either one or two bound states. For a small coupling $G$ [domain I in Fig.3(c)], the system shows a single bound state with energy slightly modified from the unperturbed value $\sigma$ and given by
\begin{equation}
E_1=\frac{\left(  \sigma/2+ \sqrt{(\sigma/2)^2-G^2-1} \right)^2+(1+G^2)^2}{(1+G^2) \left(  \sigma/2+ \sqrt{(\sigma/2)^2-G^2-1} \right) }.
\end{equation}
In the framework of the Lee model, such a state represents the "physical" particle state of the $V$ fermion with renormalized mass \cite{r24}. However, as $G$ is increased, in addition to the "physical" particle state a new bound state appears at the energy
\begin{equation}
E_2=\frac{\left(  \sigma/2- \sqrt{(\sigma/2)^2-G^2-1} \right)^2+(1+G^2)^2}{(1+G^2) \left(  \sigma/2- \sqrt{(\sigma/2)^2-G^2-1} \right) }.
\end{equation}
which is called a "ghost" [domain II in Fig.3(c)]. As discussed in Ref.\cite{r25}, the appearance of a ghost state in addition to the physical V-particle state can be detected by monitoring the temporal evolution of the occupation probability $P(t)=|c_0(t )|^2$, when the
system is initially prepared in the bare V state, i.e. for $c_n(0)=\delta_{n,0}$: the existence of the ghost state is visualized as an undamped oscillatory behavior of $P(t)$ that arises from the interference of the physical and ghost states.

The most challenging issue toward an experimental implementation of the tight-binding lattice relies on the realization of the non-Hermitian coupling $g=-iG$ of site $|0 \rangle$ with the semi-array. A few proposals have been previously suggest, based on a fast temporal modulation of complex on-site energies \cite{r25,r26}, however such methods are rather challenging in practice and, as a matter of fact, to date there are not any experimental demonstration of non-Hermitian complex couplings in tight-binding networks. A simple way to synthesize  a complex hopping, and thus the Lee Hamiltonian (29) in the ghost regime, is shown in Fig.3(d). The two terminating sites in the main lattice S are connected to an auxiliary site A with complex potential $U$. The Hamiltonian of S is given by
\begin{eqnarray}
(\mathcal{H}^{(S)})_{n,m} & =  & \kappa (\delta_{n,m+1}+\delta_{n,m-1})+ \sigma_1 \delta_{n,0}\delta_{m,0}+\sigma_2 \delta_{n,1}\delta_{m,1} \nonumber \\
& + & (\theta-\kappa)(\delta_{n,0}\delta_{m,1}+\delta_{n,1} \delta_{m,0} )
\end{eqnarray}
($n,m=0,1,2,....$), where: $\kappa$ is the hopping rate between adjacent sites, with a defective hopping rate $\theta$ between sites $|0 \rangle_S$ and $|1 \rangle_S$; and $\sigma_1$, $\sigma_2$ are the potentials at sites $|0 \rangle_S$ and $|1 \rangle_S$, respectively. The auxiliary site, with complex potential $U$, is connected to sites $|0 \rangle_S$ and $|1 \rangle_S$ via a hopping amplitude $\omega$. Following a similar procedure than the one discussed in the previous example (Sec.III.A), after elimination of the auxiliary site A the following effective energy-dependent Hamiltonian for S is found
\begin{equation}
(\mathcal{H}_{eff} )_{n,m}=\mathcal{H}^{(S)}_{n,m}+\frac{\omega^2}{E-U} \sum_{k,l=0}^{1} \delta_{n,l}\delta_{m,k}.
\end{equation}
Since we wish to simulate the Lee Hamiltonian at energies in proximity of the physical particle-V state, an approximate energy-independent Hamiltonian can be obtained from Eq.(33) by letting $E=E_0$, where $E_0$ is an energy close to either the particle or ghost state energies $E_{1,2}$. Taking for example $E=E_2$, the effective Hamiltonian (33) reduces to the Lee Hamiltonian (29) with complex coupling provided that the site potentials $U$, $\sigma_1$ and $\sigma_2$ are tuned at the values
\begin{eqnarray}
U & = & \frac{\omega^2}{\theta+iG}+E_2 \\
\sigma_1 & = & \sigma + \theta+iG \\
\sigma_2 & = & \theta + i G.
\end{eqnarray}
For example, let us consider the Lee Hamiltonian for parameter values $G/ \kappa=1.05$ and $\sigma/ \kappa=3$, i.e. inside the domain II of Fig.3(b) and corresponding to the existence of two bound states (the physical V state and the ghost state). To implement such an Hamiltonian, we assume $\omega/ \kappa=7$, $\theta / \kappa=0.2$ and tune the values of $U$, $\sigma_1$ and $\sigma_2$ according to Eqs.(34-36), namely $\sigma_1 / \kappa=3.2+1.05 i$, $\sigma_2/ \kappa=0.2+1.05i$ and $U/ \kappa \simeq 11 -45i$. To check the fidelity of the synthesized Hamiltonian, in Fig.3(e) we compare the numerically-computed evolution of the occupation probability $P(t)=|c_0(t)|^2$ with the initial condition $c_n(0)= \delta_{n,0}$, as obtained by the exact Lee Hamiltonian with complex coupling [Eq.(29)] and by the synthesized effective Hamiltonian  [Eq.(33)]. Note that the oscillatory behavior of the occupation probability, arising form the interference of the physical particle state and the ghost state, is satisfactorily reproduced by the effective Hamiltonian.

\subsection{Bound states in the continuum in a $\mathcal{PT}$-symmetric tight-binding lattice}
Bound states in the continuum (BIC) are quite anomalous bound states with energy embedded into the continuous spectrum of scattered states. In simple terms, they can be viewed as resonances of zero width. Originally predicted by Von Neumann and Wigner in certain slowly-decaying oscillating potentials \cite{r40}, they have been later found to arise from quite different mechanisms. In experiments, BICs have been predicted and observed in a  wide range of physical systems, such as  condensed-matter, electromagnetic, optical, acoustical and hydrodynamic systems \cite{r38,r41,r42,r43,r44,r45}. In particular, classical and quantum tight-binding  networks provide a fertile platform to tailor the energy spectrum and to synthesize BIC modes \cite{r38,r45,r46}. While most of previous studies on BIC states have been limited
to considering Hermitian systems, recent works have extended the idea of BIC modes to $\mathcal{PT}$-symmetric non-Hermitian photonic networks \cite{r27,r47,r48}, where they can appear either below or above the symmetry breaking threshold. While BIC states above the symmetry breaking threshold are quite common and are similar to bound states outside the continuum because they have complex energies \cite{r47}, BIC modes in the unbroken $\mathcal{PT}$ phase are less common and their synthesis requires special lattice engineering \cite{r27}. In particular, a non-Hermitian $\mathcal{PT}$-symmetric lattice sustaining one BIC mode below the symmetry breaking threshold can be synthesized following the proposal of Ref.\cite{r27}, however complex hopping rates are required. The tight-binding Hamiltonian of the lattice is given by \cite{r27}
\begin{equation}
\hat{H}= \sum_{n=-\infty}^{\infty} \left( \kappa_{n+1} |n+1 \rangle \langle n | + \kappa_n |n \rangle \langle n-1| \right)
\end{equation}
with inhomogeneous hopping amplitudes given by
\begin{equation} 
\frac{\kappa_n} { \kappa} =\left\{ 
\begin{array}{lll}
\sqrt{(n+1)/(n-1)} & n \; {\rm even} \;, \; n \neq 0 & \kappa_0=-ig \;\;\;\;\\
\sqrt{(n-2)/n} & n \; {\rm odd} \;, \; n \neq 1 & \kappa_{1}=ig  \;\;\;\;
\end{array}
\right.
\end{equation}
\begin{figure}
\includegraphics[scale=0.35]{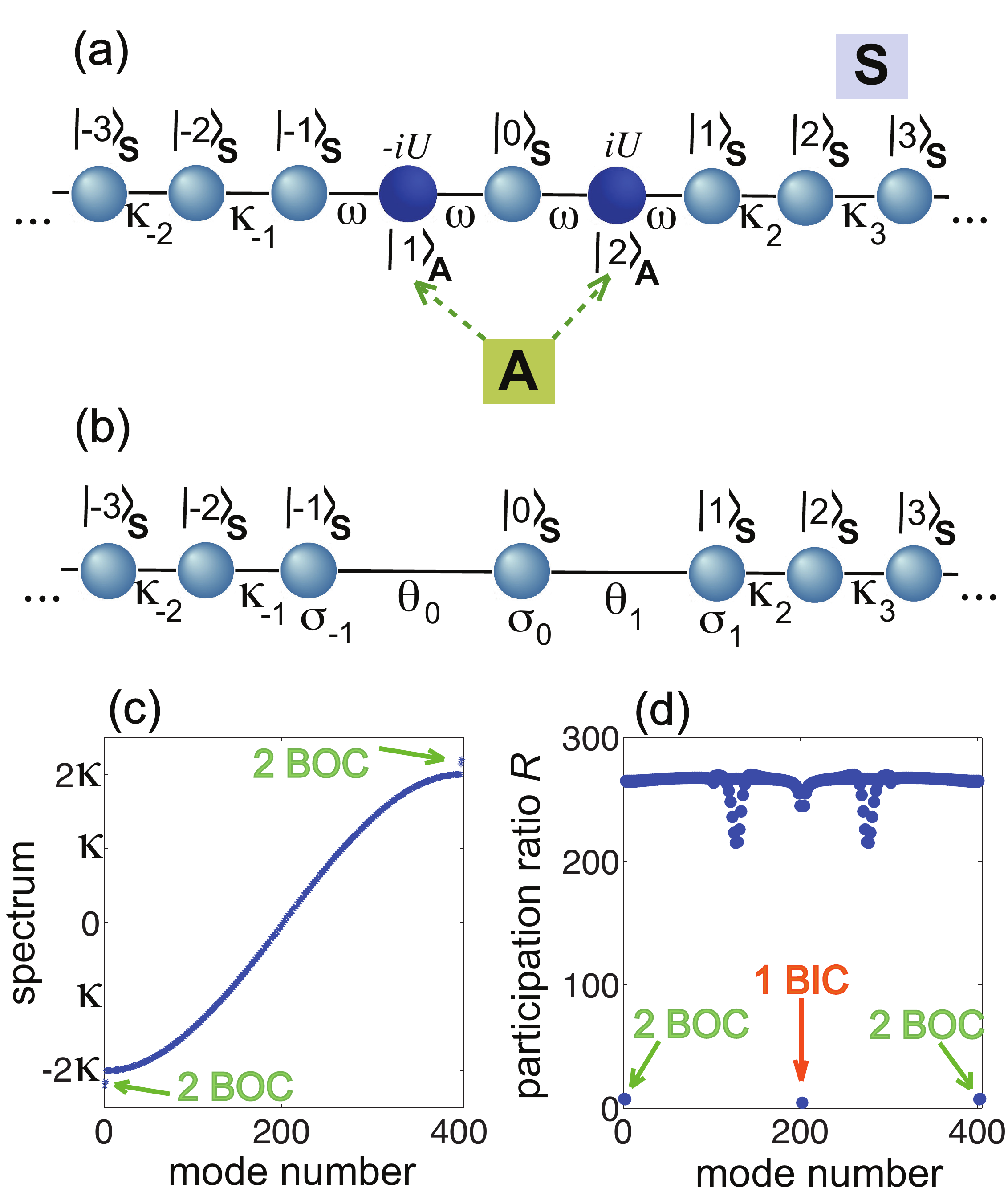}
\caption{(Color online) (a) Schematic of the $\mathcal{PT}$-symmetric tight-binding lattice, with Hermitian hopping amplitudes and imaginary potentials $\pm iU$ at the auxiliary sites $|1 \rangle_A$ and $|2 \rangle_A$,  that admits of a bound state in the continuum. The hopping rates $\kappa_n$ ($n \neq 0,1$) are defined by Eq.(37) given in the text. (b) Equivalent tight-binding lattice model obtained after elimination of the two auxiliary sites. The energy-dependent hopping amplitudes $\theta_{0,1}$ and site potentials $\sigma_{0,1,2}$ are given by $\theta_0(E)=\omega^2 / (E+iU)$, $\theta_1(E)=\omega^2 / (E-iU)$, $\sigma_{-1}(E)=\omega^2 / (E+iU)$, $\sigma_0(E)=2 E \omega^2 /(E^2+U^2)$, and $\sigma_1(E)=\omega^2 /(E-iU)$. (c) Numerically-computed energy spectrum of the tight-binding lattice of Fig.4(a) for $\omega / \kappa=1$ and $U/ \kappa=0.4$. The lattice comprises $N=403$ sites. Eigenmodes are ordered for increasing values of eigenenergy. The two arrows in the figure highlight the existence of four bound states with energies in the gap, two above and the other two below the tight-binding energy band. (d) Numerically-computed participation ratio $R$ of the lattice eigenmodes. Localized modes, corresponding to low values of $R$, are highlighted by the arrows in the figure. The central eigenmode, with energy $E_0=0$ in the middle of the allowed band, corresponds to the BIC state.}
\end{figure}
where $g>0$ is a real-valued parameter. Note that, since $\kappa_n / \kappa \rightarrow 1$ as $n \rightarrow \pm \infty$, the lattice is asymptotically homogeneous. It also satisfies the $\mathcal{PT}$ symmetry requirement  $\kappa_{-n}=\kappa_{n+1}^*$. The non-Hermitian nature of the lattice arises from the imaginary value of the hopping amplitudes $\kappa_0$ and $\kappa_1$. The energy spectrum of $\hat{H}$ is real-valued for $g \leq g_{th}=\kappa$, i.e. $\mathcal{PT}$ symmetry breaking occurs at $g_{th}=\kappa$. In the unbroken $\mathcal{PT}$ phase ($g < g_{th}$), the energy spectrum comprises, in addition to the continuous spectrum $(-2 \kappa, 2 \kappa)$ of scattered states of the asymptotic homogeneous lattice, one BIC mode with algebraic localization at the energy $E_0=0$, given by \cite{r27}
\begin{equation} 
{c}_n =\left\{ 
\begin{array}{ll}
0 & n \; {\rm odd}  \;\;\;\;\\
\kappa /g & n=0 \\
\frac{n}{|n|} \frac{i^{n+1}}{ \sqrt{n^2-1}} & n \; {\rm even} ,\; n \neq 0 \;\;\;\;
\end{array}
\right.
\end{equation}
Here we suggest a simpler $\mathcal{PT}$-symmetric tight-binding lattice, which does not require complex hopping rates and that admits of the same BIC state. It comprises a tight-binding network S and two auxiliary sites $|1\rangle_A$ and $|2\rangle_A$ in the geometrical setting of Fig.4(a). For such a system we can write
\begin{equation}
\mathcal{H}^{(S)}_{n,m}= 
\left\{
\begin{array}{ll}
\kappa_n \delta_{n,m+1}+\kappa_{n+1} \delta_{n,m-1} & {\rm both} \; n, m \neq 0 \\
0 & {\rm either} \; n,m =0
\end{array}
\right.
\end{equation}
\begin{equation}
\mathcal{H}^{(A)}= \left(
\begin{array}{cc}
-iU & 0 \\
0 & iU
\end{array}
\right)
\end{equation}
\begin{equation}
\rho_{\alpha,n}=\omega \delta_{\alpha,1} (\delta_{n,-1}+\delta_{n,0})+\omega \delta_{\alpha,2} (\delta_{n,0}+\delta_{n,1})
\end{equation}
where $\kappa_n$ (with $n \neq 0,1$) are defined by Eq.(38), and $\omega$, $U$ are real parameters. After elimination of the auxiliary sites, according to Eq.(11) the following effective energy-dependent Hamiltonian is obtained 
\begin{eqnarray}
\left( \mathcal{H}_{eff} \right)_{n,m} & = &  \mathcal{H}^{(S)}_{n,m} \nonumber \\
& + & \frac{\omega^2}{E+iU} (\delta_{n,-1}\delta_{m,-1}+\delta_{n,-1}\delta_{m,0}+\delta_{n,0}\delta_{m,-1}) \nonumber \\ 
& + & \frac{\omega^2}{E-iU}(\delta_{n,0}\delta_{m,1}+\delta_{n,1}\delta_{m,0}+\delta_{n,1}\delta_{m,1}) \nonumber \\
& + & \frac{2 E \omega^2}{E^2+ U^2} \delta_{n,0} \delta_{m,0}
\end{eqnarray}
which is illustrated in the scheme of Fig.4(b). Note that, for $E=E_0=0$, the effective Hamiltonian (43) is equivalent to the Hamiltonian (37) with $g=\omega^2 / U$, except for additional energy potentials $ \mp i \omega^2 / U$ at the odd sites $n= \pm 1$ [Fig.4(b)]. Since the BIC mode of the Hamiltonian (37) does not occupy odd sites of the lattice [see Eq.(39)], it follows that the Hamiltonian (43), i.e. the lattice depicted in Fig.4(a), has one BIC state at energy $E_0=0$ as well. It should be noted that, owing to the dependence of $\mathcal{H}_{eff}$ on the energy $E$, its energy spectrum is not equivalent to the one of the Hamiltonian (37), even thought they admit the same BIC mode at energy $E_0=0$. In particular, the lattices of Fig.4(a) sustains additional bound states in the gap, i.e., bound states outside the continuum (BOC), while the original Hamiltonian (37) does not. We numerically computed the energy spectrum of the lattice of Fig.4(a) for $\omega / \kappa=1$ and for increasing values of the on-site potential $U/ \kappa$, assuming typically $N=403$ lattice sites with reflective boundary conditions \cite{note}. The spectrum turns out to be real (unbroken $\mathcal{PT}$ phase) for $U/\kappa < \sim 0.46$. As an example, Fig.4(c) shows the numerically-computed spectrum for $U / \kappa =0.4$. The degree of localization of the eigenstate $c_n(E)$ with energy $E$ is measured by the participation ratio $R(E)$,
given by $R(E)=(\sum_n |c_n|^2)^2/ (\sum_n |c_n|^4)$. For localized
modes, $R \sim 1$ while for extended states $R \sim N$. The distribution
of $R(E)$ for the $N=403$  lattice eigenmodes is shown in Fig. 4(d). The figure clearly indicates the
existence of one BIC state with algebraic localization at energy $E_0=0$, together with four
 BOCs with exponential decay tails and with
energies outside the lattice band. The outer BOC states
have an energy $E_{1,2} \simeq \pm 2.202 \kappa$ and $E_{3,4} \simeq \pm 2.142 \kappa$.

\section{Conclusions}
In this paper we have suggested a simple method  to engineer a tight-binding quantum network by judicious coupling to an auxiliary cluster. Remarkably, the technique allows one to implement effective non-Hermitian hopping rates with only complex on-site energies and Hermitian couplings in both the network and auxiliary cluster, avoiding the use of external time-dependent control fields. As compared to other engineering methods, such as those based on inverse scattering, supersymmetry or external time-dependent control fields, the method turns out to be rather simple and flexible for a practical implementation. We have discussed three applications of the method to timely problems: the synthesis of a nearly transparent defect in an Hermitian linear lattice; the realization of the Fano-Anderson model with complex coupling; and the synthesis of a $\mathcal{PT}$-symmetric tight-binding lattice with a bound state in the continuum.

\appendix
\section{Effective Hamiltonian description in the weak coupling limit}
In this Appendix we briefly discuss the possibility to derive an energy-independent effective Hamiltonian $\hat{H}_{eff}$ for the tight-binding network S after elimination of the degree of freedoms of the auxiliary system A. To this aim, let us introduce the vectors of amplitudes $\mathbf{c}(t)=(... c_{-1},c_0,c_1,c_2, ...)^T$ and $\mathbf{a}(t)=(a_1,a_2,a_3,....)^T$ and write the coupled differential equations (6,7) in the compact form
\begin{eqnarray}
i \frac{d \mathbf{c}}{dt} & = & \mathcal{H}^{(S)} \mathbf{c}+ \rho^T \mathbf{a} \\
i \frac{d \mathbf{a}}{dt} & = & \mathcal{H}^{(A)} \mathbf{a}+ \rho \mathbf{c}. 
\end{eqnarray}
Equation (A2) can be formally integrated, yielding
\begin{equation}
\mathbf{a}(t)=-i \int_0^t d \xi \exp[i \mathcal{H}^{(A)} (\xi-t)] \rho \mathbf{c}(\xi).
\end{equation}
In writing Eq.(A3) we assumed that the auxiliary sites are not excited at initial time, i.e. $\mathbf{a}(0)=0$. Such a condition is not necessary when the auxiliary cluster A is dissipative, i.e. all eigenvalues of $\mathcal{H}^{(A)}$ have negative imaginary part. Substitution of Eq.(A3) into Eq.(A1) yields the following integro-differential equation for $\mathbf{c}(t)$ 
\begin{equation}
i \frac{d \mathbf{c}}{dt}  = \mathcal{H}^{(S)} \mathbf{c}-i   \int_0^t d \xi  \rho^T \exp[i \mathcal{H}^{(A)} (\xi-t)] \rho \mathbf{c}(\xi).
\end{equation}
The standard derivation of an effective Hamiltonian from the integro-differential equation (A4) within Markovian approximation generally requires that \cite{r29} (i) the auxiliary cluster A has a continuous spectrum, i.e. $M \rightarrow \infty$, and (ii) the S-A coupling is week, i.e. $ \rho \rightarrow 0$ (Weisskopf-Wigner approximation). For a finite number $M$ of sites in A, such a reduction can not be generally accomplished. There is, however, a special case where it can be done and that deserves to be briefly mentioned, although it has some narrow application for the purpose of network engineering. Let us assume that (i) S is Hermitian, so they the eigenvalues of $\mathcal{H}^{(S)}$ are real; (ii) A is non-Hermitian and dissipative, with all eigenvalues of $\mathcal{H}^{(A)}$ with  negative imaginary part; (iii) S-A coupling is weak, i.e. $\rho \rightarrow 0$. In this case, it is worth considering the dynamics in the interaction picture. After setting 
\begin{equation}
\mathbf{c}(t)=  \exp(-i \mathcal{H}^{(S)} t) \tilde{\mathbf{c}}(t),
\end{equation}
 form Eqs.(A4) and (A5) one obtains
 \begin{eqnarray}
i \frac{d \mathbf{\tilde{c}}}{dt}  &  = &  -i  \exp(i \mathcal{H}^{(S)} t)  \times \\
& \times & \int_0^t d \tau  \rho^T \exp (-i \mathcal{H}^{(A)} \tau) \rho  \exp[i \mathcal{H}^{(S)} (\tau- t)] \mathbf{\tilde{c}}(t-\tau). \nonumber 
\end{eqnarray}
 In the $\rho \rightarrow 0$ limit, $\tilde{\mathbf{c}}(t)$ varies slowly on time, and $\tilde{\mathbf{c}}(t-\tau)$ under the sign of integral on the right hand side of Eq.(A6) can be calculated at $\tau=0$, since $\exp(-i \mathcal{H}^{(A)} \tau) \rightarrow 0$ and $\exp(i \mathcal{H}^{(S)} \tau)$ remains limited at $\tau \rightarrow \infty$. After extending the upper integral limit on the right hand side of Eq.(A6) to $\infty$, the integro-differential equation(A6) simplifies into the following differential equation   
\begin{equation}
i \frac{d \mathbf{\tilde{c}}}{dt}  =  \exp(i \mathcal{H}^{(S)} t) \Phi \exp(-i \mathcal{H}^{(S)} t) \mathbf{\tilde{c}}(t)
\end{equation}
where we have set
\begin{equation}
\Phi \equiv -i \int_0^{\infty} d \tau  \rho^T \exp (-i \mathcal{H}^{(A)} \tau) \rho \exp (i \mathcal{H}^{(S)} \tau)
\end{equation}
In terms of the original amplitude $\mathbf{c}(t)$, using Eqs.(A5) and (A7) one finally obtains
\begin{equation}
i \frac{d \mathbf{c}}{dt}  =  \mathcal{H}_{eff} \mathbf{c}(t)
\end{equation}
where we have set
\begin{equation}
\mathcal{H}_{eff}=\mathcal{H}^{(S)}+ \Phi.
\end{equation}
Therefore, coupling with the auxiliary dissipative cluster A yields a correction to the the tight-binding Hamiltonian of the network S, given by the term $\Phi$ defined by Eq.(A8). However, since the above derivation holds in the weak coupling approximation, the correction $\Phi$ to $\mathcal{H}^{(S)}$ is  generally a small one.


\begin{thebibliography}{99}

\bibitem{r1}
D. D'Alessandro, {\it Introduction to Quantum Control and Dynamics} (Chapman and Hall/CRC, London, 2007)

\bibitem{r2}
H. Rabitz, New J. Phys. {\bf 11}, 105030 (2009).

\bibitem{r3}
P. Doria, T. Calarco, and S. Montangero, Phys. Rev. Lett. {\bf 106}, 190501 (2011).

\bibitem{r3bis}
E. Torrontegui, S. Martinez-Garaot, and J.G. Muga, Phys. Rev. A {\bf 89}, 043408 (2014).

\bibitem{r4}
S. Bose, Phys. Rev. Lett. {\bf 91},  207901 (2003); S. Bose, Contemp. Phys. {\bf 48}, 13 (2007).

\bibitem{r5}
M. Christandl, N. Datta, A. Ekert, and A. J. Landahl, Phys. Rev. Lett. {\bf 92}, 187902 (2004).

\bibitem{r6}
A.D. Greentree, S.J. Devitt, and L.C.L. Hollenberg, Phys. Rev. A {\bf 73}, 032319 (2006).

\bibitem{r7}
P. Rebentrost, I. Serban, T. Schulte-Herbr\"{u}ggen, and F. K. Wilhelm, Phys. Rev. Lett. {\bf 102}, 090401 (2009).

\bibitem{r7b}
A.  Kay, Int. J. Quantum Inf. {\bf 8}, 641 (2010).

\bibitem{r8}
G.M. Nikolopoulos and I. Jex, {\it  Quantum State Transfer and Network Engineering} (Springer, Berlin, 2014).

\bibitem{r9}
 D. Jaksch and P. Zoller, New J. Phys. {\bf 5}, 56 (2003);
 F. Dreisow, A. Szameit, M. Heinrich, T. Pertsch, S. Nolte, A. T\"{u}nnermann, and S. Longhi, Phys. Rev. Lett. {\bf 101}, 143602 (2008);
 S. Longhi, Phys. Rev. Lett. {\bf 101}, 193902 (2008);
M. Aidelsburger, M. Atala, S. Nascimbene, S. Trotzky, Y.A.Chen, and I. Bloch, Phys. Rev. Lett.
{\bf 107} , 255301 (2011); M. Aidelsburger, M. Atala, M. Lohse, J. T. Barreiro,
B. Paredes, and I. Bloch, Phys. Rev. Lett. {\bf 111}, 185301 (2013).

\bibitem{r9b}
M. M\"{u}ller, S. Diehl, G. Pupillo, and P. Zoller, Adv. Atom. Mol. Opt. Phys. {\bf 61},1 (2012).

\bibitem{r9c}
A. Ajoy and P. Cappellaro, Phys. Rev. Lett. {\bf 110}, 220503 (2013); D. Hayes, S.T. Flammia, and M.J. Biercuk, New J. Phys. {\bf 16}, 083027 (2014).

\bibitem{r10}
T.D. Stanescu, V. Galitski, J.Y. Vaishnav, C.W. Clark, and S. Das Sarma, Phys. Rev. A {\bf 79}, 053639 (2009). 

\bibitem{r11}
Y.J. Lin, R.L. Compton, K. Jimenez-Garcia, J.V. Porto, and I.B.  Spielman, Nature {\bf 462}, 628 (2009).

\bibitem{r12}
 N.H. Lindner, G. Refael, V. Galitski, Nature Phys. {\bf 7}, 495 (2011).
 
 \bibitem{r13}
 A. Gomez-Leon and G. Platero, Phys. Rev. Lett. {\bf 110}, 200403 (2013).
 
 \bibitem{r14}
 T. Tanamoto, K. Ono, Yu-Xi Liu, and F. Nori, Sci. Rep. {\bf 5}, 10076 (2015).

\bibitem{r15}
N. Goldman and J. Dalibard, Phys. Rev. X {\bf 4}, 031027 (2014).

\bibitem{r16}
G.M. Nikolopoulos, D. Petrosyan, and P.J. Lambropoulos, J. Phys.: Condens. Matter {\bf 16},
4991 (2004); M. Bellec, M. Georgios, M. Nikolopoulos, and S. Tzortzakis,
Opt. Lett. {\bf 37}, 4504 (2012).

\bibitem{r17}
N. Moiseyev, {\it Non-Hermitian Quantum Mechanics} (Cambridge University Press, Cambridge, 2011)

\bibitem{r18}
C. Bender, Rep. Prog. Phys. {\bf 70}, 947 (2007). 

\bibitem{r18b}
L. Jin and Z. Song, Phys. Rev. A {\bf 84}, 042116 (2011). 

\bibitem{r18c}
H. Zhong, W. Hai, G. Lu, and Z. Li, Phys. Rev. A {\bf 84}, 013410  (2011).

\bibitem{r19}
X.Z. Zhang and Z. Song, Ann. Phys. {\bf 339}, 109 (2013);
X. Lian, H. Zhong , Q. Xie, X. Zhou, Y. Wu, and W. Liao, Eur. Phys. J. D {\bf 68}, 189 (2014);
S. Lin, X. Z. Zhang, and Z. Song Phys. Rev. A {\bf 92}, 012117 (2015);  J. Gong and Q.-H. Wang
Phys. Rev. A {\bf 91}, 042135 (2015); T.E. Lee and Y.N. Joglekar,
Phys. Rev. A {\bf 92}, 042103 (2015).

\bibitem{r20}
B.T. Torosov, G. Della Valle, and S. Longhi,
Phys. Rev. A {\bf 87}, 052502 (2013); G. Della Valle and S. Longhi, Phys. Rev. A {\bf 87}, 022119 (2013); 
X. Luo, J. Huang, H. Zhong, X. Qin, Q. Xie, Y.S. Kivshar, and C. Lee,
Phys. Rev. Lett. {\bf 110}, 243902 (2013); S. Longhi and G. Della Valle, Phys. Rev. A {\bf 87}, 052116 (2013); 
S. Ibanez and J. G. Muga, Phys. Rev. A {\bf 89}, 033403 (2014); 
J. D'Ambroise, B.A. Malomed, and P.G. Kevrekidis, 
Chaos {\bf 24}, 023136 (2014);  C. Yuce, Eur. Phys. J. D {\bf 69}, 184 (2015).

\bibitem{r21}
M. Znojil,  	J. Phys.: Conf. Ser. {\bf 624},  012011  (2015).

\bibitem{r22}
S. Zhang, Z. Ye, Y. Wang, Y. Park, G. Bartal, M. Mrejen, X. Yin, and X. Zhang,
Phys. Rev. Lett. {\bf 109}, 193902 (2012); P. Ginzburg, F.J. Rodríguez-Fortuno, A. Martinez, and A.V. Zayats, 
Nano Lett. {\bf 12}, 6309 (2012); G. Castaldi, S. Savoia, V. Galdi, A. Al\'{u}, and N. Engheta,
Phys. Rev. Lett. {\bf 110}, 173901 (2013); C.M. Gentry and M.A. Popovic, Opt. Lett. {\bf 39}, 4136 (2014); 
L. Feng, X. Zhu, S. Yang, H. Zhu, P. Zhang, X. Yin, Y. Wang, and X. Zhang, Opt. Express {\bf 22}, 1760 (2014); 
H. Hodaei, M.-A. Miri, M. Heinrich, D.N. Christodoulides, and M. Khajavikhan, Science {\bf 346}, 975 (2014). 

\bibitem{r23}
X. Z. Zhang, L. Jin, and Z. Song, Phys. Rev. A {\bf 87}, 042118 (2013)

\bibitem{r24}
C.M. Bender, S.F. Brandt, J.-H. Chen, and Q. Wang, Phys. Rev. D {\bf 71}, 025014 (2005).

\bibitem{r25}
S. Longhi and G. Della Valle, Phys. Rev. A {\bf 85}, 012112 (2012). 

\bibitem{r26}
S. Longhi, Phys. Rev. A {\bf 82}, 032111 (2010).

\bibitem{r27}
S. Longhi, Opt. Lett. {\bf 39}, 1697 (2014); S. Longhi and G. Della Valle, Phys. Rev. A {\bf 89}, 052132 (2014).

\bibitem{r28}
S. Longhi, Phys. Rev. A {\bf 92}, 042116 (2015).

\bibitem{r29}
L. Fonda, G. C. Ghirardi, and A. Rimini, Rep. Prog. Phys. {\bf 41},
587 (1978); P.L. Knight, M.A. Lauder, and B.J. Dalton, Phys. Rep. {\bf 190}, 1 (1990); H. Nakazato, M. Namiki, and S. Pascazio, Int. J. Mod. Phys. B
{\bf 10}, 247 (1996); S. Longhi, J. Mod. Opt. {\bf 56}, 729 (2009).

\bibitem{r30}
M.M. Sternheim and J.F. Walker, Phys. Rev. C {\bf 6}, 114 (1972).

\bibitem{r31}
 I. Rotter, Rep. Prog. Phys. 54, 635 (1991); V.V. Sokolov and V.G. Zelevinsky,
Ann. Phys. {\bf 216}, 323 (1992); G.G. Giusteri, F. Mattiotti, and G.L. Celardo,
Phys. Rev. B {\bf 91}, 094301 (2015).

\bibitem{r32}
A.A. Sukhorukov, Opt. Lett. {\bf 35}, 989 (2010);
S. Longhi and G. Della Valle, Phys. Rev. B {\bf 84}, 193105 (2011); 
U.A. Khawaja and A.A. Sukhorukov, Opt. Lett. {\bf 40},  2719 (2015).

\bibitem{r33}
A. Szameit, F. Dreisow, M. Heinrich, S. Nolte, and A.A. Sukhorukov,
Phys. Rev. Lett. {\bf 106}, 193903 (2011).

\bibitem{r34}
S. Longhi, Opt. Lett. {\bf 40}, 463 (2015).

\bibitem{r35}
U. Fano, Phys. Rev. {\bf 124}, 1866 (1961); P.W. Anderson, Phys. Rev. {\bf 164}, 41 (1961).

\bibitem{r36}
 G.D. Mahan, {\it Many-Particle Physics} (New York, Plenum Press, 1990), pp. 272-285; M. Cini, {\it Topics and Methods in Condensed-Matter Theory}
(Springer, Heidelberg, 2007), Chap. 5, pp. 81-89.

\bibitem{r37}
 K.O. Friedrichs, Commun. Pure Appl. Math. {\bf 1}, 361 (1948);
T.D. Lee, Phys. Rev. {\bf 95}, 1329 (1954); I. Prigogine, Phys. Rep. {\bf 219}, 93 (1992).

\bibitem{r38}
P. A. Orellana, F. Dominguez-Adame, I. Gomez, and M. L.
Ladron de Guevara, Phys. Rev. B {\bf 67}, 085321 (2003); P. Orellana
and F. Dominguez-Adame, Phys. Stat. Sol. A {\bf 203}, 1178 (2006);
A.V. Malyshev, P.A.Orellana, and F. Dominguez-Adame, Phys.
Rev. B {\bf 74}, 033308 (2006); I. Rotter and A. F. Sadreev, Phys. Rev. E {\bf 69}, 066201 (2004);
H. Nakamura, N. Hatano, S. Garmon, and T. Petrosky, Phys.
Rev. Lett. {\bf 99}, 210404 (2007); S. Garmon, H. Nakamura,
N. Hatano, and T. Petrosky, Phys. Rev. B {\bf 80}, 115318 (2009); S. Longhi,Eur. Phys. J. B {\bf 57}, 45 (2007); 
S. Longhi, Phys. Rev. B {\bf 75}, 184306 (2007); R. Farchioni, G.Grosso, and G.P. Parravicini, Eur. Phys. J. B {\bf 84}, 227 (2011).

\bibitem{r39}
 A.E. Miroshnichenko, S. Flach, and Y.S. Kivshar,  Rev. Mod. Phys. {\bf 82}, 2257 (2010).
 
 \bibitem{r39b}
 W.H. Louisell, {\it Radiation and Noise in Quantum Electronics} (Mc-Graw-Hill, New York, 1964), sec. 7.6.
 
 \bibitem{r39c}
 R.J. Glauber, Annals of the New York Academy of Sciences {\bf 480}, 336 (1986).
 
\bibitem{r40}
J. von Neumann and E. Wigner, Z. Phys. 30, 465 (1929).

\bibitem{r41}
W. Koch,  J. Sound Vib. {\bf 88}, 233 (1983); M.D. Groves,  Math.
Method. Appl. Sci. {\bf 21}, 479 (1998); E. Davies and L. Parnovski, Q. J. Mech. Appl. Math. {\bf 51}, 477 (1998); 
S. Hein, W. Koch, and L. Nannen, J. Fluid Mech. {\bf 692}, 257 (2012).

\bibitem{r42}
M. Callan, C.M. Linton, and D.V. Evans,  J. Fluid Mech. {\bf 229}, 51 (1991); P.J. Cobelli, V. Pagneux, A. Maurel, and P. Petitjeans, EPL {\bf 88}, 20006 (2009).

\bibitem{r43}
J.U. N\"{o}ckel, Phys. Rev. B {\bf 46}, 15348 (1992); E.N. Bulgakov, P. Exner, K.N. Pichugin, A.F. Sadreev,
Phys. Rev. B {\bf 66}, 155109 (2002); S. Longhi, J. Opt. Soc. Am. B {\bf 11}, 1098 (1994); L.S. Cederbaum, R.S. Friedman, V.M. Ryaboy, and N. Moiseyev, Phys. Rev. Lett. {\bf 90}, 013001 (2003);
 M.L. Ladron de Guevara, F. Claro, P.A. Orellana, Phys. Rev. B {\bf 67}, 195335 (2003); I. Rotter, A.F. Sadreev, Phys. Rev. E 71, 046204 (2005); M.L. Ladron de Guevara, and P.A. Orellana, Phys. Rev. B {\bf 73}, 205303 (2006);  W.-J. Gong, X.-Y. Sui, Y. Wang, G.-D. Yu, and X.-H. Chen, Nanoscale Res. Lett. {\bf 8}, 330 (2013).	

\bibitem{r44}
D.C. Marinica, A.G. Borisov, and S.V. Shabanov, Phys. Rev. Lett. {\bf 100}, 183902 (2008); 
E.N. Bulgakov and A.F. Sadreev, Phys. Rev. B {\bf 78}, 075105 (2008); 
M. I. Molina, A. E. Miroshnichenko, and Y. S. Kivshar, Phys.
Rev. Lett. {\bf 108}, 070401 (2012);
C. W. Hsu, B. Zhen, J. Lee, S.-L. Chua, S. G. Johnson, J. D.
Joannopoulos, and M. Soljacic, Nature {\bf 499}, 188
(2013); Y. Yang, C. Peng, Y. Liang, Z. Li, and S. Noda,
Phys. Rev. Lett. {\bf 113}, 037401  (2014); B. Zhen, C.W. Hsu, L. Lu, A.D. Stone, and M. Soljacic,
Phys. Rev. Lett. {\bf 113}, 257401 (2014).

\bibitem{r45}
F. Dreisow, A. Szameit, M. Heinrich, R. Keil, S. Nolte, A. T\"{u}nnermann, and S. Longhi, Opt. Lett. {\bf 34}, 2405 (2009); 
Y. Plotnik, O. Peleg, F. Dreisow, M. Heinrich, S. Nolte, A.
Szameit, and M. Segev, Phys. Rev. Lett. {\bf 107}, 183901 (2011); G. Corrielli, G. Della Valle, A. Crespi, R. Osellame, and S.
Longhi, Phys. Rev. Lett. {\bf 111}, 220403 (2013); S. Weimann, Y. Xu, R. Keil, A. E. Miroshnichenko, A. Tunnermann, S. Nolte,
A. A. Sukhorukov, A. Szameit, and Y. S. Kivshar, Phys. Rev.
Lett. {\bf 111}, 240403 (2013).

\bibitem{r46}
K.-K. Voo and C.S. Chu, Phys.Rev. B {\bf 74}, 155306 (2006);
S. SreeRanjani. A.K. Kapoor, and P.K. Panigrahi, AIP Conf. Proc. {\bf 864}, 236 (2006); 
J. M. Zhang, D. Braak, and M. Kollar, Phys. Rev. Lett. {\bf 109}, 116405 (2012);
 S. Longhi and G. Della Valle, Sci. Rep. {\bf 3}, 2219 (2013);  
 J. Mur-Petit and R.A. Molina, Phys. Rev. B {\bf 90}, 035434 (2014);
 C.-L. Zou, J.-M. Cui, F.-W. Sun, X. Xiong, X.-B. Zou, Z.-F. Han, and G.-C. Guo, 
 Laser \& Photon. Rev. {\bf 9},  114 (2015). 
  	
\bibitem{r47}
A. Regensburger, M.-A. Miri, C. Bersch, J. N\"{a}ger, G.
Onishchukov, D. N. Christodoulides, and U. Peschel, Phys.
Rev. Lett. {\bf 110}, 223902 (2013).

\bibitem{r48}
Y.N. Joglekar, D.D. Scott, and A. Saxena,
Phys. Rev. A {\bf 90}, 032108  (2014);
M.I. Molina and Y.S. Kivshar, Studies Apple. Math. {\bf 133}  337 (2014);
S. Garmon, M. Gianfreda, and N. Hatano,
Phys. Rev. A {\bf 92}, 022125 (2015).

\bibitem{note}
We checked that, for a sufficiently large number of sites $N$, lattice truncation does not introduce boundary effects (such as surface states) and the energy spectrum is almost insensitive to the precise value of $N$.



\end{thebibliography}
\end{document}